\documentclass[aps,pra,onecolumn,12pt,showpacs,showkeys,superscriptaddress]{revtex4-1}
\usepackage{mathrsfs}
\usepackage{amsmath}
\usepackage{amsfonts}
\usepackage{amssymb}
\usepackage{amsthm}
\usepackage{graphicx}
\usepackage{color}
\usepackage{graphics}
\usepackage{epsfig}
\usepackage{ulem}
\usepackage{subfigure}%
\usepackage{epstopdf}
\usepackage{lineno}
\providecommand{\U}[1]{\protect\rule{.1in}{.1in}}
\begin{document}
\title{Anomalous magnon Nernst effect of topological
magnonic materials}

\author{X S Wang}
\affiliation{School of Microelectronics and Solid-State Electronics,
University of Electronic Science and Technology of China, Chengdu,
Sichuan 610054, China}
\affiliation{Department of Physics, The Hong Kong University of
Science and Technology, Clear Water Bay, Kowloon, Hong Kong}
\author{X R Wang}
\email[Corresponding author:]{phxwan@ust.hk}
\affiliation{Department of Physics, The Hong Kong University of
Science and Technology, Clear Water Bay, Kowloon, Hong Kong}
\affiliation{HKUST Shenzhen Research Institute, Shenzhen 518057, China}

\begin{abstract}
The magnon transport driven by thermal gradient in a
perpendicularly magnetized honeycomb lattice is studied.
The system with the nearest-neighbor pseudodipolar interaction and
the next-nearest-neighbor Dzyaloshinskii-Moriya interaction (DMI)
has various topologically nontrivial phases. When an in-plane thermal
gradient is applied, a transverse in-plane magnon current is generated.
This phenomenon is termed as the anomalous magnon Nernst effect that
closely resembles the anomalous Nernst effect for an electronic system.
The anomalous magnon Nernst coefficient and its sign are determined
by the magnon Berry curvatures distribution in the momentum space
and magnon populations in the magnon bands.
We predict a temperature-induced sign reversal in anomalous
magnon Nernst effect under certain conditions.
\end{abstract}
\maketitle
\section{Introduction}
Spintronics is about generation, detection and manipulation of
spin degree of freedom of particles. Most early studies focused
on the electron spins \cite{DasSarma}. However, an electric current
normally accompanies an electron spin current and consumes much
energy, leading to a Joule heating. The Joule heating becomes the
critical problem in nano electronics and spintronics although many
efforts have been made.  Recently, magnon spintronics, or magnonics
in which magnons are spin carriers, attracts much attention
because of its fundamental interest \cite{xiansi,hubin} and its
lower energy consumption in comparison with that of electron
spintronics \cite{book1,magnonics1,magnonics2}.

Nernst effect commonly refers to the generation of a
transverse voltage/current by a thermal gradient in an
electronic system under a perpendicular magnetic field.
In a ferromagnetic metal and in the absence of an external magnetic
field, a thermal gradient can generate a transverse charge current
or voltage proportional to the vector product of the thermal
gradient and the magnetization in the linear response region.
This is the anomalous Nernst effect, the thermal electric
manifestation of the anomalous Hall effect \cite{ANE}.
It is natural to ask whether there is a similar effect for magnons.
Moving magnons experience gyroscopic forces because of nonzero
Berry curvature of a magnetic system although magnons are charge
neutral quasiparticles that do not have the Lorentz force.
As a result, a transverse magnon current is generated when
magnons are driven by a longitudinal force such as a thermal
gradient in the absence of a magnetic field which is termed
as the anomalous magnon Nernst effect (AMNE).
In this paper, we focus on a perpendicularly magnetized
honeycomb lattice with the nearest-neighbor pseudodipolar
interaction and the next-nearest-neighbor Dzyaloshinskii-Moriya
interaction (DMI), whose magnon bands can be topologically
nontrivial with various topological phases \cite{ours}.
We investigate the magnon transport of this system in the presence
of a thermal gradient using the semiclassical equations of motion
of magnons and the Boltzmann equation in linear transport regime.
We found that the system has topologically nontrivial magnon bands.
The system changes from one topologically nontrivial phase to another
as the DMI strength varies. The AMNE coefficient depends on temperature
nonmonotonically. It starts from 0 at 0 K and goes back to 0 at high
temperature limit with a maximum at an intermediate temperature.
The nonmonotonical temperature-dependence of AMNE is due to non-trivial
Berry curvature distribution of a given band in the momentum space
and thermally activated magnon population in the bands.
In certain parameter space, there is a sign reversal of the AMNE at
low temperature because the magnon Berry curvature near the band bottom
at $\Gamma$ point has small non-zero values of the opposite sign as
those near band top at K and K$^\prime$ points with a much bigger value.
In the presence of staggered anisotropy on A, B sublattices, the system
can also be topologically trivial, and the K and K$^\prime$ valleys
contribute opposite transverse magnon currents due to the opposite
Berry curvatures. However, the total transverse magnon current does
not vanish. The boundary that AMNE coefficient changes its sign is
also determined numerically.

\begin{figure}[htb]
\centering
\includegraphics[width=8.5cm]{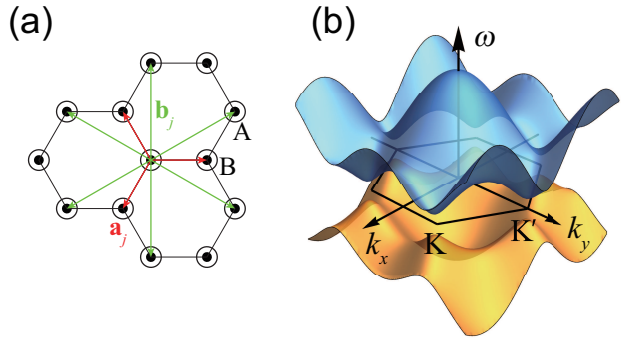}\\
\caption{(a) Schematic illustration of a perpendicularly
magnetized honeycomb lattice. The red and green arrows denote
the nearest-neighbor and the next-nearest-neighbor vectors, respectively.
(b) Magnon spectrum $\omega(\mathbf{k})$ of an infinite system for
$K=10J$, $F=5J$, and $D=\Delta=0$.
The Brillouin zone is indicated by the black hexagon.
}
\label{system}
\end{figure}

\section{Model and Results}
We consider classical magnetic moments on a honeycomb
lattice in the $xy$ plane as illustrated in Figure 1(a),
and the Hamiltonian is
\begin{multline}
\mathcal{H}=-\frac{J}{2}\sum_{\left\langle i,j \right\rangle} \mathbf{m}_i\cdot
\mathbf{m}_j-\frac{F}{2}\sum_{\left\langle i,j \right\rangle}
(\mathbf{m}_i\cdot \mathbf{e}_{ij})(\mathbf{m}_j\cdot \mathbf{e}_{ij})
-D\sum_{\langle\langle i,j\rangle\rangle}\nu_{ij}
\hat{\mathbf{z}}\cdot\left(\mathbf{m}_i\times\mathbf{m}_j\right)-\sum_{i}\frac{K_i}{2}m_{iz}^2,
\label{Hami}
\end{multline}
where the first term is the nearest-neighbor ferromagnetic Heisenberg
exchange interaction ($J>0$). The second and third terms arise from
the spin-orbit coupling (SOC) \cite{pseudo,DMI}. $\mathbf{e}_{ij}$ is
the unit vector pointing from site $i$ to $j$. $F$ is the strength
of the nearest-neighbor pseudodipolar interaction, which is the
second-order effect of the SOC [The nearest-neighbor
Dzyaloshinskii-Moriya interaction (DMI) would be the first-order
effect of SOC if it exists, but it vanishes because the center of
the A-B bond is an inversion center of the honeycomb lattice].
The next-nearest-neighbor DMI measured by $D$ is in general no zero.
$\nu_{ij}=\frac{2}{\sqrt{3}}\hat{\mathbf{z}}\cdot(\mathbf{e}_{li}
\times\mathbf{e}_{lj})=\pm 1$, where $l$ is the nearest neighbor
site of $i$ and $j$. The last term is the sublattice-dependent
anisotropy whose easy-axis is along the $z$ direction with anisotropy
coefficients of $K_i=K+\Delta$ for $i\in$ A and $K-\Delta$ for $i\in$ B.
$\mathbf{m}_i$ is the unit vector of the magnetic moment at site $i$.
The spin dynamics is governed by the Landau-Lifshitz-Gilbert (LLG)
equation \cite{LLG,ours},
\begin{equation}
\frac{\mathrm{d}\mathbf{m}_i}{\mathrm{d} t}=-\gamma\mathbf{m}\times \mathbf{H}^
\text{eff}_i+\alpha \mathbf{m}_i\times \frac{\mathrm{d}\mathbf{m}_i}{\mathrm{d} t},
\label{LLG}
\end{equation}
where $\gamma$ is the gyromagnetic ratio and $\alpha$ is the Gilbert
damping constant. $\mathbf{H}^\text{eff}_i=\frac{\partial\mathcal{H}}
{\mu_0\mu\partial\mathbf{m}_i}$ is the effective field at site $i$.
The lattice constant $a$ and $J$ are used as the length unit
and the energy unit out of five parameters in \eqref{Hami}.
The magnetic field and time are in the units of $\sqrt{J\mu_0/a^3}$
and $\sqrt{a^3\mu_0/(J\gamma^2)}$, respectively, where $\mu_0$ is
the vacuum permeability. When the anisotropy is sufficiently large,
spins are perpendicularly magnetized \cite{split}.
To obtain the spin wave spectrum, we linearize the LLG equation
following the standard procedures \cite{ours}.
The spin wave spectrum and wavefunctions are obtained by solving the
eigenvalue problem $gH(\mathbf{k})\psi_n=\omega_n(\mathbf{k})\psi_n$,
where $H(\mathbf{k})$ is a $4\times 4$ Hermitian matrix
\begin{equation}H=
\left(\begin{matrix}
M_\mathrm{A}^{+} & 0 & -\ell(\mathbf{k}) & -g_{+}(\mathbf{k})\\
0 & M_\mathrm{A}^{-} & -g_{-}(\mathbf{k}) & -\ell(\mathbf{k})\\
-\ell^{*}(\mathbf{k})  & g_{-}^{*}(\mathbf{k}) & M_\mathrm{B}^{-} & 0\\
-g_{+}^{*}(\mathbf{k}) & \ell^{*}(\mathbf{k}) & 0 & M_\mathrm{B}^{+}
\end{matrix}\right),
\label{eigen}
\end{equation}
with $\ell(\mathbf{k})=\left(J+\frac{F}{2}\right)
\sum_{j=1,2,3}e^{i\mathbf{k}\cdot\mathbf{a}_j}$, $g_{\pm}
(\mathbf{k})=\frac{F}{2}\sum_{j=1,2,3}e^{\pm2i\theta{j}}
e^{i\mathbf{k}\cdot\mathbf{a}_j}$ ($\theta_{j}$ is the angle
between $\mathbf{a}_j$ and $x$ direction).
$M_\mathrm{A}^{\pm}=M+\Delta\pm d(\mathbf{k})$ and
$M_\mathrm{B}^{\pm}=M-\Delta\pm d(\mathbf{k})$ with $M=K+3J$ and
$d(\mathbf{k})=2D\sum_{j=1,3,5}\sin(\mathbf{k}\cdot\mathbf{b}_j)$.
$g=\sigma_0\bigotimes\sigma_3$ (with $\sigma_0$ being the
$2\times2$ identity matrix and $\sigma_3$ the Pauli matrix).
$\psi_n$ is the $n$th eigenvector of eigen-frequency $\omega_n$,
satisfying the generalized orthogonality $\psi_i^\dagger g
\psi_j=\delta_{ij}$. At K and K$^\prime$, the frequencies of
the two magnon bands are respectively,
\begin{eqnarray}
\omega_{1}^\mathrm{K(K^\prime)}=M-3\sqrt{3}D+(-)\Delta,\label{gap1}\\
\omega_{2}^\mathrm{K(K^\prime)}=\sqrt{(M+3\sqrt{3}D)^2-\frac{9}{4}F^2}-(+)\Delta,
\label{gap2}\end{eqnarray}
where ``$(+)$" and ``$(-)$" on the right hand side are for K$^\prime$.
The magnon band for $K=10J$, $F=5J$ and $D=\Delta=0$ is shown in
Figure 1(b), which has a direct gap of $\Delta_g=M-\sqrt{M^2-9F^2/4}$
at both K and K$^\prime$ (valleys for the upper band and peaks for
the lower band). The direct gap at the valleys can close and reopen as
$D$ and $\Delta$ varies, resulting in topological phase transitions.
The Berry curvature $\boldsymbol{\Omega}_{n}$ of $n$th band and the
corresponding Chern number $\mathcal{C}_n$ can be calculated
by using a gauge-invariant formula \cite{Shindou},
\begin{gather}
\boldsymbol{\Omega}_{n}=i\nabla_\mathbf{k}\times\left(\psi_n^\dagger g
\nabla_\mathbf{k}\psi_n\right);\\
\mathcal{C}_n=\frac{1}{2\pi}\iint_{\mathbf{k}\in \mathrm{BZ}}\Omega_n d^2\mathbf{k},
\end{gather}
where the integration is over the Brillouin zone (BZ), and
$\Omega_n=\boldsymbol{\Omega}_n\cdot\hat{\mathbf{z}}$ is the $z$
component of the Berry curvature that is given by a gauge-invariant
formula similar to that in electronic systems \cite{chern}
\begin{equation}
\boldsymbol{\Omega}_{n}=i\mathrm{Tr}\left[P_n\left(\frac{\partial P_n}
{\partial k_x}\frac{\partial P_n}{\partial k_y}-\frac{\partial P_n}{\partial
k_y}\frac{\partial P_n}{\partial k_x}\right)\right]\hat{\mathbf{z}},
\end{equation}
where $P_n$ is the projection matrix
of the $n$th band defined as $P_n=\psi_n  \psi_n^\dagger g$.

\begin{figure}
\centering
\includegraphics[width=8.cm]{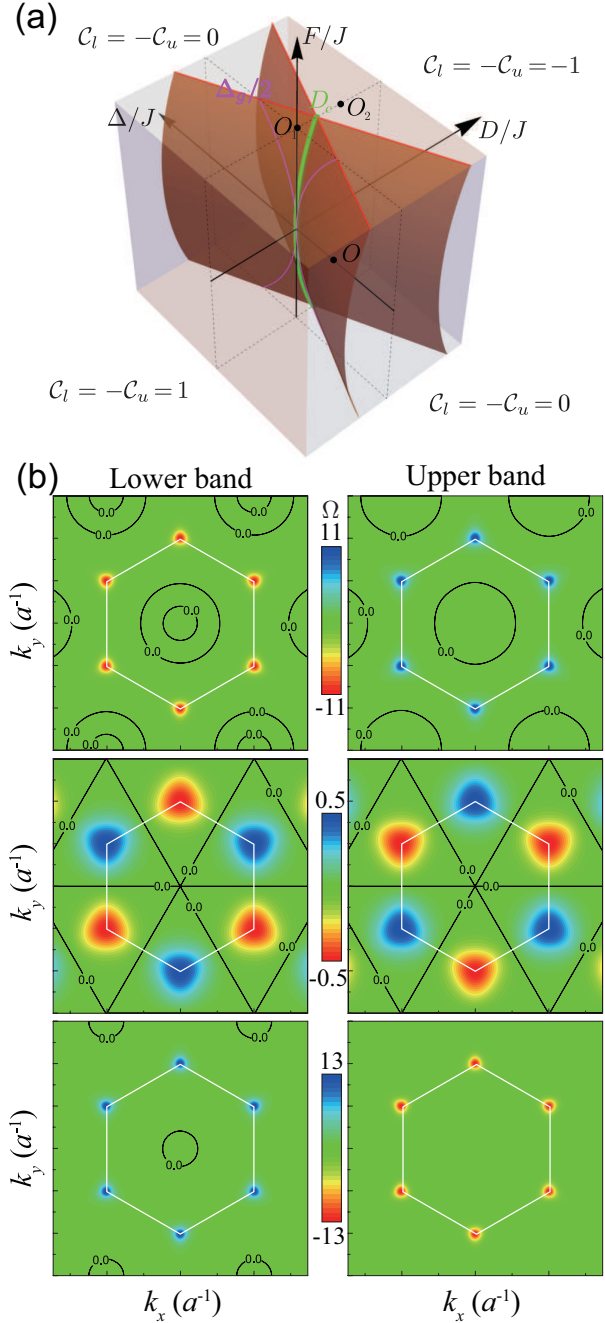}\\
\caption{(a) Phase diagram in $D/J-\Delta/J-F/J$ space for $K=10J$.
Phases are classified by the Chern numbers of the upper and lower
magnon bands. Three phases of $C_l=-C_u=1$; $C_l=-C_u=-1$; and
$C_l=C_u=0$ are separated by two orange boundary surfaces.
The green line, $D=\frac{\sqrt{3}F^2}{16M}$ and
$\Delta=0$, is the intersection of the two boundary surfaces.
The magenta lines, $\Delta=\pm\Delta_g/2$ and $D=0$, are the
intersections of the boundary surfaces with the $D=0$ plane.
$O_1$ ($F=5J$, $D=\Delta=0$), $O_2$ ($F=5J$, $D=0.4J$, $\Delta=0$),
and $O$ ($F=D=0$, $\Delta=-1.5J$) are 3 representative points
in topologically nontrivial phase of $C_l=-C_u=-1$; $C_l=-C_u=1$,
and topologically trivial phase $C_l=C_u=0$, respectively.
(b) The $z$ component of Berry curvature
$\Omega=\boldsymbol{\Omega}\cdot\hat{\mathbf{z}}$ for $O_1$, $O$,
$O_2$ (from top to bottom). The left panel is for the lower magnon
band and the right panel for the upper magnon band.
The color bars are shown at the middle. The contour line of
$\Omega=0$ is shown by the dashed circles.
The white hexagons is the fisrt Brillouin zone.
}
\label{berry}
\end{figure}
Figure 2(a) is the phase diagram in $D/J-\Delta/J-F/J$ space for $K=10J$.
The various topological phases are classified by Chern numbers
$\mathcal{C}_l$ and $\mathcal{C}_u$ of lower and upper magnon bands.
$\mathcal{C}_l+\mathcal{C}_u=0$ satisfies the ``zero sum rule"
\cite{chern,book_niu}. The magnon band Chern number change its value
when magnon band gap closes and reopens at valley K or K$^\prime$.
Thus, the band gap closing at K or K$^\prime$ defines two phase
boundary surfaces of $\omega_1^{K\prime}=\omega_2^{K\prime}$ and
$\omega_1^K=\omega_2^K$ (See Eqs.  \eqref{gap1} and \eqref{gap2}).
For convenience, we define
\begin{equation}
\Delta_c=\frac{1}{2}\left[\sqrt{(M+3\sqrt{3}D)^2-\frac{9}{4}F^2}-(M-3\sqrt{3}D)\right],
\end{equation}
and two phase boundary surfaces are $\Delta=\pm \Delta_c$, denoted as
the orange surfaces. They divide the whole space into four regions.
In the region of $\Delta_c<0$ and $\Delta_c<\Delta<-\Delta_c$,
$C_u$ is $1$. The density plot of $\Omega$ for $F=5J$, $\Delta=D=0$
($O_1$ in Fig. 2(a)) is shown in the top panel of Fig. 2(b).
Interestingly, the lower band has two contour curves of
$\Omega=\Omega_l=0$ around $\Gamma$ denoted by black dash lines.
The two contour curves divide the first Brillouin zone into three parts.
$\Omega$ is slightly positive inside the inner contour curve
around $\Gamma$ for the lower band as shown in the top left panel.
Between two contour curves, $\Omega$ is slightly negative.
$\Omega$ is positive outside the outer contour curve as shown
in the top left panel of Fig. 2(b), but significant non-zero
$\Omega$ occurs only around K and K$^\prime$.
In the region of $\Delta_c>0$ and $-\Delta_c<\Delta<\Delta_c$, the
upper magnon band has Chern number $-1$. The bottom panel of Fig. 2(b)
is the density plot of $\Omega$ of lower (left panel) and upper
(right panel) bands for a representative point of $F=5J$, $\Delta=0$,
$D=0.4J$ ($O_2$ in Fig. 2(a)) in this topologically nontrivial phase.
The lower band has only one contour curve of $\Omega=0$ (black dash
curve) around $\Gamma$ that divides the first Brillouin zone into
two parts. Inside the contour curve, $\Omega$ is slightly positive
as shown in the bottom left  panel of Fig. 2(b).
It is negative outside the contour curve with significant non-zero
value around K and K$^\prime$. The system is in topologically trivial
phase for both lower and upper bands in the other two regions.
$\Omega$ around K and K$^\prime$ valleys have opposite sign so that
the Chern numbers are 0 for both bands. We consider $O$ in Fig. 2(a)
($F=D=0$, $\Delta=1.5J$) as a representative point in the phase.
The middle panel of Fig. 2(b) shows the density plot of $\Omega$
at $O$ for the two bands. Indeed, Berry curvatures $\Omega$ at K
and K$^\prime$ have opposite value, and Chern numbers are zeros.
For $\Delta=0$, the band gaps at K and K$^\prime$ close and reopen
at the same time and the Chern number of the upper band changes
from $-1$ to $+1$ if we tune the DMI crossing the line of $D=\frac
{\sqrt{3}F^2}{16M}$ and $\Delta=0$ [the green line in Figure 2(a)].
The system changes from one topologically nontrivial phase to another.
The features of the phase diagram discussed above preserves as long
as system ground state is the perpendicular ferromagnetic state.

Let us consider the magnon transport in an infinite system.
Apply a thermal gradient along $x$ direction, the motion of a magnon
wavepacket is governed by the semiclassical equations \cite{Niu,Murakami},
\begin{eqnarray}
\dot{\mathbf{r}}=\frac{1}{\hbar}\frac{\partial \varepsilon}{\partial
\mathbf{k}}-\dot{\mathbf{k}}\times\boldsymbol{\Omega};
\label{c1}\\
\dot{\mathbf{k}}=\frac{1}{\hbar}\mathbf{F}=-\frac{1}{\hbar}\frac
{\partial \varepsilon}{\partial \mathbf{r}}+\frac{q}{\hbar}
\dot{\mathbf{r}}\times\mathbf{B}, \label{c2}
\end{eqnarray}
Where $\varepsilon(\mathbf{r},\mathbf{k})=\hbar\omega(\mathbf{k})+\phi(
\mathbf{r})$ is the energy of the magnon with $\phi(\mathbf{r})$ being
the potential energy, and $\mathbf{F}$ is the total force on the magnon.
$q$ is the charge of the particle and $q=0$ for a magnon.
In the presence of a thermal gradient, the Boltzmann equation of the magnon is
\begin{equation}
\dot{\mathbf{r}}\cdot \frac{\partial f}{\partial \mathbf{r}}=
-\frac{f-f_0}{\tau}\equiv-\frac{f_1}{\tau}, \label{B.E.}
\end{equation}
where $f(\mathbf{r},\mathbf{k})$ is the magnon distribution function.
$f_0=1/(e^{\beta\hbar\omega}-1)$ is the Bose-Einstein distribution of
zero chemical potential at local temperature $T$ [$\beta=(k_BT)^{-1}$].
$\tau$ is magnon relaxation time. $f_1$ is the deviation of the
distribution function from its equilibrium values.
In the linear response regime where the thermal gradient is small,
Eq. \eqref{B.E.} can be written as
\begin{equation}
\dot{\mathbf{r}}\cdot \frac{\partial f_0}{\partial \mathbf{r}}=-\frac{f_1}{\tau}.
\end{equation}
One can prove the following identity,
\begin{equation}
\dot{\mathbf{r}}\cdot \frac{\partial f_0}{\partial \mathbf{r}}
=\left(-\frac{\hbar\omega}{T}\nabla T\right)\cdot\frac{\partial f_0}{\hbar\partial \mathbf{k}}.
\end{equation}
Substituting Eq. (14) into the left hand side of Eq. (13), it yields
\begin{equation}
\left(-\frac{\hbar\omega}{T}\nabla T\right)\cdot\frac{\partial f_0}{\hbar\partial \mathbf{k}}=-\frac{f_1}{\tau}.
\end{equation}
Thus, one can identify a thermal force $\mathbf{F}_T=\left(-
\frac{\hbar\omega}{T}\nabla T\right)$ proportional to the
magnon frequency and the thermal gradient \cite{thermalforce}.
Insert  \eqref{c2} into  \eqref{c1} with $\mathbf{F}=\mathbf{F}_T$,
we obtain
\begin{equation}
\dot{\mathbf{r}}=\frac{\partial \omega}{\partial \mathbf{k}}+\frac{\omega}{T}
\nabla T\times\boldsymbol{\Omega}.
\end{equation}
The magnon current density is given by $\mathbf{j}_m=
\sum_{n,\mathbf{k}}\left[\dot{\mathbf{r}}f(n,\mathbf{k})\right]$,
where the summation is over all magnon states.
Keep terms linear in the thermal gradient and convert the
summation to integration, we have
\begin{equation}
\mathbf{j}_m=\tensor{\boldsymbol{\kappa}}(-k_B\nabla T),
\end{equation}
where the longitudinal heat conductance $\kappa_{xx}$ and the
anomalous Nernst coefficient $\kappa_{xy}$ are
\begin{eqnarray}
\kappa_{xx}=\frac{\tau}{(2\pi)^2}\sum_{n}\iint \beta\left(
\frac{\partial \omega_n}{\partial k_x}\right)^2
\rho\left(\beta\hbar\omega_n \right) d^2\mathbf{k},\label{kxx}\\
\kappa_{xy}=\frac{1}{(2\pi)^2}\sum_{n}\iint\beta\omega_n
\Omega_nf_0\left(\beta\hbar\omega_n\right)d^2\mathbf{k}\label{kxy},
\end{eqnarray}
where $\rho(x)=\frac{xe^x}{(e^x-1)^2}$, $f_0(x)=\frac{1}{e^x-1}$,
and $n=1,2$ labels the lower and upper magnon bands.
Figure 3(a) shows the temperature dependence of $\kappa_{xx}$
and $\kappa_{xy}$ in two different topologically-nontrivial
phases specified by $O_1$ and $O_2$ in Fig. 2(a).
In order to have a quantitative feeling about the results, we use
$\mathrm{Sr_2IrO_4}$ parameters of $a=0.55$ nm \cite{para},
$J=19.6\mu_0\mu_B^2/a^3$ \cite{pseudo}, and $\gamma=2.21 \times10^5$ rad/s/(A/m) in all the following discussions.
The longitudinal heat conductance $\kappa_{xx}$ is always positive
as expected from thermodynamic laws that the magnons move from
the hot side to the cold side. Eq. \eqref{kxy} says that the AMNE
coefficient is determined by the Berry curvature distribution in the
momentum space and the magnon equilibrium distribution function.
Since magnon number in the lower band is bigger than that in the higher
band according to the Bose-Einstein distribution, the sign of AMNE
coefficient is always determined by the Berry curvature of the lower
magnon band. At very low temperature, only the magnons near$\Gamma$
point [band bottom (top) of the lower (upper) band) are excited.
The sign of AMNE coefficient is determined by $\Omega$ around
$\Gamma$, and its value is small because Berry curvature $\Omega$
is very close to zero, if not exactly zero, and the magnon number
is also small there. At a higher temperature when the magnon number
near K and K$^\prime$ points [band top (bottom) of the lower (lower)
band] are large enough and dominate the AMNE due to significant
non-zero values of the Berry curvature only near there.
At even higher temperature when equal-partition theorem become true
so that $f_0\approx k_BT/(\hbar\omega)$, the AMNE coefficient is
close to zero because $\kappa_{xy}$ is approximately proportional
to $(\mathcal{C}_u+\mathcal{C}_l)=0$ \cite{chern,book_niu},
i.e. the contributions from two bands cancel with each other.

The general behavior of AMNE coefficient $\kappa_{xy}$ mentioned
above can be illustrated by two representative points in two
distinct topologically nontrivial phases of $C_l=-1$ (for $O_1$)
and $C_l=1$ (for $O_2$). For $O_1$ whose Berry curvature distribution
is given in the top panel of Fig. 2(b), $\kappa_{xy}$ is always
positive, a transverse magnon current along $\mathbf{m}_0\times
(-\nabla T)$, because $\Omega$ are positive near both $\Gamma$
and K ($K^\prime$) points. For $O_2$, at very low temperatures when
the magnon number around K and K$^\prime$ are negligible and only
the magnons near $\Gamma$ point are excited, $\kappa_{xy}$ decreases
and becomes more and more negative initially with the increase
of temperature because $\Omega$ is negative near $\Gamma$ point.
However, when magnons near K and K$^\prime$ points are excited, $\kappa_{xy}$ starts to increase with temperature, and becomes
postive after an intermediate temperature because $\Omega$ has
large positive values near K and K$^\prime$. Thus, in this phase
the sign of the AMNE coefficient reverses at the intermediate
temperature. The numerical results of $\kappa_{xx}$ and $\kappa_{xy}$
at higher temperature are shown in the inset of Figure 3(a).
The longitudinal heat conductance $\kappa_{xx}$ saturates at
high temperature. AMNE coefficient $\kappa_{xy}$ at $O_1$
($O_2$) increases from 0 to a maximum positive (negative)
value as the temperature increases, and then gradually go
back to 0 when magnons in the upper band are thermally excited.
This indicates that there is an optimal temperature for the
maximal AMNE coefficient. If this temperature does not exceed
the Curie temperature, it should be used for the largest AMNE.

\begin{figure}
\centering
\includegraphics[width=8.5cm]{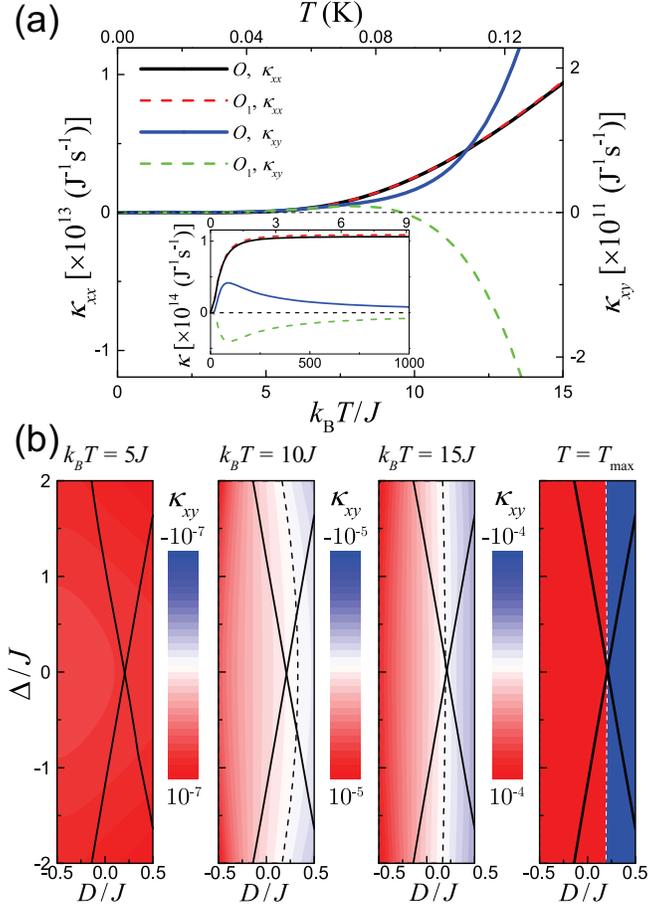}\\
\caption{(a) The longitudinal magnon conductance $\kappa_{xx}$
(left axis) and AMNE coefficient $\kappa_{xy}$ (right axis) for
parameters at $O_1$ and $O_2$. The inset shows the high-temperature
values of the same quantities. (b) (Panels 1 to 3) The density plots
of $\kappa_{xy}$ [in units of $\mathrm{eV^{-1}s^{-1}}$] in $D/J$-
$\Delta/J$ plane for $K=10J$ and $F=5J$ at different temperatures.
The black solid lines are topological phase boundaries, and the
black dashed lines are the contour lines of $\kappa_{xy}=0$. WHY COLOR BARS are negative on top and po on bot.? What are the use of black lines?
(Panel 4) The sign of the maximum value of $\kappa_{xy}$.
Red region is for positive $\kappa_{xy}$ and blue region is for
negative $\kappa_{xy}$.
The dashed line is the contour curve of $\kappa_{xy}=0$.
}
\label{results}
\end{figure}

In the topologically trivial phase, the Berry curvatures $\Omega$
Of the same band has opposite values near K and K$^\prime$ points.
Thus the contributions to AMNE from different valleys cancel each
other, and the net transverse magnon current can be in either
direction, depending on the parameters. Figure 3(b) is the density
plots of $\kappa_{xy}$ as a function of $D/J$ and $\Delta/J$ at
different temperatures (for $K=10J$ and $F=5J$).
Because of the featured distribution of Berry curvature near $\Gamma$
discussed above, the sign change of $\kappa_{xy}$ happens at larger
$D$ at lower temperatures, and is different to the topological
phase boundaries as shown by the black solid lines.
However, the sign change of $\kappa_{xy}$ is closely related to the
topological phase transition, as shown in the last panel of Figure 3(b).
The sign change of $\kappa_{xy}$ coincides with the topological phase
transition line of $D=\frac{\sqrt{3}F^2}{16M}$ and $\Delta=0$.
Tuning the DMI can drive the system from one topologically
nontrivial phase to another at $\Delta=0$.
The sign change of $\kappa_{xy}$ at the maximum point changes at
the same time due to the sign-reversal of Berry curvatures.
This also means for the parameters
of negative $\kappa_{xy}$ in the last panel, there is a
temperature-induced sign reversal of $\kappa_{xy}$.

In the above discussions, we studied the magnon Nernst effect, a
transverse magnon current generated by a longitudinal thermal gradient.
Similar to electronic systems, there are other related effects,
such as a transverse magnon current induced by a longitudinal
chemical potential gradient (magnon Hall effect and anomalous magnon
Hall effect), and a transverse magnon heat current induced by a
longitudinal chemical potential gradient (magnon Peltier effect).
These effects can be investigated in the same way as what
have done here for the same Berry curvature physics.
Similar topological phase
transitions and sign-reversal of AMNE was also predicted in pyrochlore
lattices \cite{mook}. In the calculation of thermal transport coefficients,
the thermal energy $k_BT$ is allowed to be much higher than $J$.
In real materials, the temperature is limited by the Curie temperature
that is order of $J/k_B$. For example, $J=20$ meV ($2.5\times10^5 \mu_0\mu^2/a^3$)
and the Curie temperature is about 240 K \cite{exp2012}
for $\mathrm{Sr_2IrO_4}$. The sign-reversal temperature is
$9.5J/k_B$ as shown in Figure 3(a). Thus, the temperature is much smaller than the
sign-reversal temperature in this
case so the AMNE coefficient should be always positive.
The reason why the Berry curvature near $\Gamma$ point has opposite
sign, and the factors that affect the Berry curvature distribution
are still open questions.

\section{Conclusion}
In conclusion, we studied the thermal magnon transport of
perpendicularly magnetized honeycomb lattice with the nearest-neighbor
pseudodipolar interaction and the next-nearest-neighbor DMI.
We show that the system has various topological nontrivial phases.
Due to the nontrivial Berry curvature, a transverse magnon current
appears when a thermal gradient is applied, resulting in an
anomalous Magnon Nernst effect. The sign of the anomalous Magnon
Nernst effect is reversed by tuning DMI and temperature.

\section*{Acknowledgements}
This work was supported by National Natural Science Foundation of
China (Grant No. 11374249) and Hong Kong RGC (Grant No. 16300117 and
16301816). X.S.W acknowledge support from UESTC and China
Postdoctoral Science Foundation (Grant No. 2017M612932).

\end{document}